\documentstyle[12pt]{article}
\newcounter{appendixc}
\newcounter{subappendixc}[appendixc]
\newcounter{subsubappendixc}[subappendixc]

\renewcommand{\appendix}[1] {\vspace*{0.6cm}
        \refstepcounter{appendixc}
        \setcounter{figure}{0}
        \setcounter{table}{0}
        \setcounter{equation}{0}
        \renewcommand{\thefigure}{\Alph{appendixc}.\arabic{figure}}
        \renewcommand{\thetable}{\Alph{appendixc}.\arabic{table}}
        \renewcommand{\theappendixc}{\Alph{appendixc}}
        \renewcommand{\theequation}{\Alph{appendixc}.\arabic{equation}}
        \noindent{\bf Appendix \theappendixc #1}\par\vspace*{0.4cm}}

\begin{document}
\begin{titlepage}
\begin{flushright}
NUHEP-TH-96-4
\end{flushright}
\vspace{0.1in}
\begin{center}
{\Large Yukawa Corrections to Single Top Quark Production \\
         at the Fermilab Tevatron in the Two-Higgs-Doublet
         Models}
\vspace{.3in}

     Chong Sheng Li{\footnote {On leave from Department of Physics,
                           Peking University, China

                ~~Electric address: csli@nuhep.phys.nwu.edu

                ~~(After December: csli@svr.bimp.pku.edu.cn)}},
     Robert J. Oakes {\footnote{ Electric address: oakes@fnal.gov}}
and  Jin Min Yang{\footnote  {On leave from Department of Physics,
                              Henan Normal University, China

                              ~~Electric address: jmyang@nuhep.phys.nwu.edu

                              ~~(After December:  yangjm@itp.ac.cn)}}

\vspace{.2in}

     Department of Physics and Astronomy, Northwestern University,\\
     Evanston, Illinois 60208-3112, USA

\end{center}
\vspace{.6in}

\begin{footnotesize}
\begin{center}\begin{minipage}{5in}
\baselineskip=0.25in
\begin{center} ABSTRACT\end{center}
                    
  We calculate Yukawa corrections of order $\alpha_{ew} M_t^2/M_W^2$ to single
 top quark production via $q \bar q'\rightarrow t \bar b$ at
 the Fermilab Tevatron in the two-Higgs-doublet models. In our calculation
 we also keep the terms proportional to $ M_b^2 \tan^2 \beta$
 since their effects may become rather important for
 large $\tan\beta$.  The corrections can amount to more 
than a 15\% reduction in the production cross section relative to the tree 
level result
in the general two-Higgs-doublet model,
and a 10\% enhancement in the minimal supersymmetric model,
which might be observable at a high-luminosity Tevatron.

\end{minipage}\end{center}
\end{footnotesize}
\vspace{.8in}

PACS number: 14.80Dq; 12.38Bx; 14.80.Gt

\end{titlepage}
\eject
\baselineskip=0.25in

\begin{center}1. Introduction \end{center}

  The top-quark physics has become a very active research area since      
the top quark was discovered by the CDF and D0 Collaborations at the
Fermilab Tevatron[1]. It is timely to focus  attention on 
directly investigating the properties of the
top quark, especially its production mechanisms. 
With the increase in the number of top quark events at the Tevatron, the
experimental errors are expected to be further reduced. With the next Tevatron
run at $\sqrt s=2.0$ TeV, one can expect about twenty times as much data as 
exist now. Thus, the comparison between the observed top quark
 production properties and more precise theoretical calculations will
be an inportant probe for the possible existence of new physics.
At the Tevatron top quarks are produced primarily via two independet 
mechanisms: The dominant production mechanism is the QCD pair production process
$q \bar q \rightarrow t \bar t$[2]. Single top production via 
$W$-gluon fusion subprocess $g+W\rightarrow t\bar b$ [3]
and the subprocess $q \bar q' \rightarrow t \bar b$ [4] are also important.
These latter processes involve the electroweak interaction and, therefore, can 
probe the electroweak sector of the theory, in contrast to the QCD pair 
production mechanism. A recent analysis[5] of the process 
$q \bar q'\rightarrow t\bar b$ showed that it is potentially observable at the
Tevatron with 2-3 $fb^{-1}$ of integrated luminosity. This process
probes the top quark with a timelike $W$ boson, $q^2 > (M_t+M_b)^2$, while
the $W$-gluon fusion process involves a spacelike $W$ boson, $q^2<0$, and 
these processes are therefore complementary.
Moreover, in the Standard Model (SM), the process  
$q\bar q' \rightarrow t\bar b$ can be reliably 
predicted and the theoretical uncertainty in the cross section is only about
a few percent due to QCD corrections[6].
Although
the statistical error in the measured cross section for this process at the
Tevatron will be about $\pm$30\% [5], a high-luminosity Tevatron would allow
a measurement of the cross section with a statistical uncertainty of about
6\%[6]. At this level of experimental accuracy a calculation of the 
radiative corrections is necessary. 
In Ref.[6] the QCD and Yukawa corrections to single
top quark production $q \bar q'\rightarrow t \bar b$ have been
calculated in the SM. While the QCD corrections were found to be quite large, 
the Yukawa
corrections were found to be negligible. Since the SM weak corrections
are expected
to be comparable to the Yukawa corrections, they too should be negligible.
Beyond the SM, the Yukawa corrections might be greatly enhanced, since
more Higgs bosons with stronger couplings to top or bottom quarks 
are involed in some new physics models.
Once the top quark mass is known precisely, these effects could be used as
an indirect test for new physics beyond the SM; for example, 
the two-Higgs-doublet model(2HDM) and the minimal supersymmetric model(MSSM)[7].
At least, the data could be used to place restrictions on these models.
Therefore, it is worthwhile to investigate 
single top quark production via $q \bar q'\rightarrow t\bar b$ in these models.
In this paper we present the calculation of the Yukawa
corrections of order $\alpha_{ew} M_t^2/M_W^2$ to single top production at
the Fermilab Tevatron in both the 2HDM and the MSSM.
These corrections arise from the virtual effects of the third family 
(top and bottom) of quarks,
neutral and charged Higgs bosons, and neutral and charged Goldstone bosons.
We note that our calculations 
can be easily extended to the pseudo-Goldstone boson (PGB) 
corrections in technicolor models[8] by
substituting the virtual PGB's in the technicolor
models for the virtual Higgs bosons in the 2HDM and  MSSM.

\begin{center} 2. Calculations \end{center}

  The tree-level Feynman diagram for single top quark production via
$q \bar q'\rightarrow t \bar b$ is shown in Fig.1(a).
  The Yukawa corrections of order $\alpha_{ew} M_t^2/M_W^2$ to the process
$q \bar q'\rightarrow t \bar b$ arise from the Feynman diagrams shown
in Figs.1b-1h. In our calculations, we used dimensional regalarization to 
control
all the ultraviolet divergences in the virtual loop corrections and we adopted
the on-mass -shell renormalization scheme[9].We also kept all the terms
proportional to the product $M_b\tan\beta$ in the charged Higgs couplings 
to the third family of quarks since
these effects may become rather important for large $\tan\beta$. 
We used the 't Hooft-Feynman gauge for the propagators of virtual $W$ boson
and Goldstone bosons.
Including the
$O(\alpha_{ew} M_t^2/M_W^2)$ Yukawa corrections, the renormalized amplitude for
$q\bar q'\rightarrow t\bar b$ can be written as
\begin{equation}
M_{ren}=M_0 + \delta M^{vertex}
\end{equation}
where $M_0$ is  the tree-level matrix element,
and $\delta M^{vertex}$ represents the $O(\alpha_{ew} M_t^2/M_W^2)$ Yukawa
corrections arising from the self-energy diagrams Figs.1b-1e and vertex
diagrams Figs.1f-1h. These are given by
\begin{equation}
 M_0= i \frac{g^2}{2} \frac{1}{\hat s-M_W^2}
  \bar v(p_2) \gamma_{\mu} P_L u(p_1)  \bar u(p_3) \gamma^{\mu} P_L v(p_4)
\end{equation}
and
\begin{equation}
\delta M^{vertex}= i \frac{g^2}{2} \frac{1}{\hat s-M_W^2}
  \bar v(p_2) \gamma_{\mu} P_L u(p_1)
  \bar u(p_3) [\gamma^{\mu} P_L \delta_1+\gamma^{\mu} P_R \delta_2
               +\bar \Lambda^{\mu}_L P_L+\bar \Lambda^{\mu}_R P_R ] v(p_4)
\end{equation}
where 
\begin{eqnarray}
 \delta_1&=&(\frac{1}{2}\delta Z^t_L+\frac{1}{2}\delta Z^b_L)_{finite}
          +f^L_1,\\
 \delta_2&=&f^R_1,\\
 \bar \Lambda^{\mu}_L&=&p^{\mu}_3 f^L_2+p^{\mu}_4 f^L_3,\\
 \bar \Lambda^{\mu}_R&=&p^{\mu}_3 f^R_2+p^{\mu}_4 f^R_3.
\end{eqnarray}
Here $p_1$ and $p_2$ denote the momentum of the incoming quarks $q$ and $\bar q'$,
while $p_3$ and $p_4$ are used for the outgoing $t$ and
$\bar b$ quarks, and $\hat{s}$ is the center-of-mass energy of the
subprocess.
$\delta Z^t_L$ and $\delta Z^b_L$ are the wave-function renormalization
constants, and $f^{L,R}_{i}$ are form factors which are presented in
Appendix A.

  The renormalized differential cross section of the subprocess is 
\begin{equation}
\frac{d\hat{\sigma}}{d\cos\theta}=\frac{\hat{s}-M_t^2}{32\pi\hat s^2}
\overline{\sum} \vert M_{ren}\vert^2,
\end{equation}
where $\theta$ is the angle between the top quark and incoming quark.
Integrating this subprocess  differential cross section over $\cos\theta$ one finds
\begin{equation}
\hat{\sigma}=\hat{\sigma}_0+\Delta \hat{\sigma}
\end{equation}
where
\begin{equation}
\hat{\sigma}_0=\frac{g^4}{128\pi}\frac{\hat{s}-M^2_t}{\hat{s}^2(\hat{s}-
M^2_W)^2}[\frac{2}{3}(\hat{s}-M_t^2)^2
 +(\hat{s}-M^2_t)(M^2_t+M^2_b) + 2M^2_tM^2_b]
\end{equation}
is the tree-level result and the correction is
\begin{eqnarray}
\Delta \hat{\sigma}&=&\frac{g^4}{64\pi}\frac{\hat{s}-M^2_t}{\hat{s}^2(\hat{s}
-M^2_W)^2}\left \{[\frac{2}{3}(\hat{s}-M^2_t)^2+(\hat{s}-M^2_t)(M^2_t+M^2_b)
\right.\nonumber\\
& &+2M^2_tM^2_b][\delta _1+\frac{1}{2}(M_tf_2^{L}-M_bf_3^{R})] +[\frac{1}{6}
(\hat{s}-M^2_t)^2+M^2_b(\hat{s}-M_t^2)+M_b^4]M_tf_3^L\nonumber\\
& &-[\frac{1}{6}(\hat{s}-M_t^2)^2+M^2_t(\hat{s}-M^2_t)+M^4_t]M_bf_2^R
+\frac{1}{2}\hat{s}(\hat{s}-M_t^2)(M_bf_3^R-M_tf^L_2)\nonumber\\
& & \left. +\hat s M_tM_b(M_tf^R_2-M_bf^L_3)
+2\hat{s}M_tM_b\delta_2\right \}.
\end{eqnarray}
The total hadronic cross section for the production of single-top-quark via $q\bar q'$
can be written in the form
\begin{equation}
\sigma (s)=\sum_{i,j}\int dx_1 dx_2 \hat\sigma_{ij}(x_1x_2s, M_t^2,
\mu^2)[f^A_i(x_1,\mu)f^B_j(x_2, \mu)+(A\leftrightarrow B)],
\end{equation}
where
\begin{eqnarray}
s&=&(P_1+P_2)^2,\\
\hat{s}&=&x_1x_2s,\\
p_1&=&x_1P_1,
\end{eqnarray}
and 
\begin{equation}
p_2=x_2P_2.
\end{equation}
Here $A$ and $B$ denote the incident hadrons and $P_1$ and $P_2$ are their
four-momenta, while $i, j$ are the initial partons and $x_1$ and $x_2$ are their
longitudinal momentum fractions. The functions $f^A_i$ and $f^B_j$ are the 
usual parton distributions. In our
numerical calculations, we have used the CTEQ3L parton distribution functions
for the tree level cross section, and CTEQ3M parton distribution functions[10]
for the $O(\alpha_{ew} M^2_t/M^2_W)$ Yukawa corrections, as in ref.[6], to 
facilitate comparison. 
There is no Yukawa correction to parton distribution functions as
pointed out in Ref.[11].
We will also compare our calculations
with those calculated using the MRSG parton distribution
functions[12] below. Finally, introducing the convenient variable 
$\tau =x_1x_2$, 
and changing independent variables, the total corss section
becomes
\begin{equation}
\sigma(s)=\sum_{i,j}\int^1_{\tau_0}\frac{d\tau}{\tau}(\frac{1}{s}
\frac{dL_{ij}}{d\tau})(\hat s \hat \sigma_{ij})
\end{equation}
where $\tau_0=(M_t+M_b)^2/s$. The quantity $dL_{ij}/d\tau$ is the parton
luminosity, which is defined to be
\begin{equation}
\frac{dL_{ij}}{d\tau}=\int^1_{\tau} \frac{dx_1}{x_1}[f^A_i(x_1,\mu)
f^B_j(\tau/x_1,\mu)+(A\leftrightarrow B)]
\end{equation}
\vspace{1cm}

\begin{center}3. Numerical results and discussions\end{center}

In the following we present some numerical results for
the Yukawa corrections to the total cross section for single
 top quark production via $q \bar q'\rightarrow t \bar b$ at
 the Fermilab Tevatron with $\sqrt s=2$ TeV.
 In our numerical calculations we chose $M_Z=91.188 GeV, M_W=80.33 GeV,
M_b=5 GeV,\alpha_{ew}=1/128$, and $\mu=\sqrt {\hat s}$. 
The Higgs masses 
$M_H,M_h,M_A,M_{H^+}$ and parameters $\alpha,\beta$ are not 
constrained in the general Two-Higgs-Doublet Model, but in the Minimal
SUSY Model, relations[13] among these parameters are required by 
supersymmetry, leaving only two parameters free;  e.g., $M_{H^+}$ and
$\tan\beta$. Also,
in the MSSM the charged Higgs mass is heavier than the $W$  mass due to
the relation $M_{H^+}^2=M_W^2+M_A^2$.
The experimental lower limit for the charged Higgs mass is 44.1GeV[14],
independent of the additional parameters $\alpha$ and $\beta$.
 In our calculations we will use $M_{H^+}=600GeV$ for 2HDM and 
$M_{H^+}>100GeV$ for MSSM.  
The upper bound of $\tan\beta$; viz, $\tan\beta<0.52 GeV^{-1}
 M_{H^+}$, has been determined from data on $B\rightarrow \tau \nu X$[15]. 
The lower limits on $\tan\beta$ are $\tan\beta>0.6$ from
perturbative bounds [16] and $\tan\beta>0.25$ (for $M_t=175$GeV) 
from perturbative unitarity[16]. We will, therefore, limit the value of 
$\tan\beta$
to be in the range of 0.25 to 30.

Figure 2 shows the relative correction 
 $\Delta\sigma/\sigma_0$  as a function of 
$M_H$ using the CTEQ3L parton distributions for the tree-level cross section
$\sigma_0$ and the CTEQ3M 
parton distributions for the correction $\Delta\sigma$, as in ref.[6].  
The solid curve corresponds to the 2HDM
assuming $M_h=M_H$ and $M_t=175 GeV, M_{H^+}=M_A=600GeV$ and $\tan\beta=0.25$.
The dotted curve corresponds to the SM for $M_t=175$ GeV.
For $M_h=M_H$ the corrections are independent of $\alpha$ in the 2HDM.
The corrections in the SM are negeligbly small, in agreement with
Ref.[6]. However, in the 2HDM, the corrections can reduce the cross section 
by more than $-10$\% for $M_h=M_H<100$GeV, and for $M_h=M_H=50$GeV they can
be as large as $-20$\%. 

Figure 3 shows the relative correction $\Delta\sigma/\sigma_0$ in the MSSM, 
assuming $\tan\beta=0.25$.
Since the corrections are not sensitive to $M_{H^+}$, for $M_{H^+}>400$GeV,
we only present the results for $M_{H^+}$ in the range 100GeV to 400 GeV.
In Fig.3 the solid curve corresponds to $M_t=175$ GeV, again using 
CTEQ3L distributions for $\sigma_0$ and CTEQ3M distributions for $\Delta\sigma$.
The dotted curve corresponds to $M_t=175$ GeV but using the MRSG parton 
distributions.
The dashed curve corresponds to $M_t=200$ GeV and using 
CTEQ3L for $\sigma_0$ and CTEQ3M for $\Delta\sigma$.
For a light charged Higgs, the corrections can be quite significant.
For $M_{H^+}=100$ GeV the corrections reach 9\% for $M_t=175$ GeV
and 13\% for $M_t=200$ GeV. 
The curve for $M_t=175$GeV has a peak at $M_{H^+}=170$ GeV
and  the curve for $M_t=200$GeV has a peak at $M_{H^+}=195$ GeV,
which is due to the fact that  $M_b=5$ GeV
 and the threshold for open top decay into a charged Higgs plus a bottom
is crossed in these regions. If we change the top quark
mass, we found that this region is also shifted correspondingly,
which provides a check on our calculations, especially of the treatment of the
threshold.
From Fig.3 we see that the difference between 
the results obtained using CTEQ3 distributions and using MRSG distributions 
is negeligibly small.
We also found that the results using MRS(A') distributions [12] are almost 
the same as the MRSG results, and thus
we did not present them. 

In both Fig.2 and Fig.3, we used the minimal value (0.25)
for $\tan\beta$. When $\tan\beta$ becomes larger, the corrections may
drop rapidly since the dominant effects arise from the terms 
$\sim \alpha_{ew} \frac{M_t^2}{M_W^2 \tan^2\beta}$.  
In Fig.4 we present the dependence of the relative correction, 
$\Delta\sigma/\sigma_0$, 
on the value of $\tan\beta$ using CTEQ3L for $\sigma_0$ and CTEQ3M 
for $\Delta\sigma$. 
The solid curve corresponds to the 2HDM
assuming $M_t=175$GeV, $M_{H^+}=M_A=600$GeV and $M_H=M_h=65$GeV.
The dotted curve corresponds to the MSSM assuming $M_t=175$GeV 
and $M_{H^+}=100$GeV.
The corrections are only significant for small  $\tan\beta$
and are quite sensitive to $\tan\beta$ for  $\tan\beta<5$.

Since the cross section for single top production can be reliably
predicted in the SM [6] and 
the statistical error in the measurment of the cross section  
will be about 6\% at a high-luminosity Tevatron[6], these corrections 
may be observable; at least, interesting new constraints on these models 
can be established.

Note that in the MSSM, besides these Yukawa corrections arising from the 
Higgs sector,  the  supersymmetric (SUSY) corrections
due to super particles (sparticles) should also be taken into account[17].
The dominant virtual effects of sparticles arise from supersymmetric QCD
corrections of order $\alpha_s$ and the supersymmetric electroweak 
correction of order $\alpha_{ew} M_t^2/M_W^2$ which arise from loops of
charginos and neutralinos, the supersymmetry partners of Higgs and vector 
bosons.
However, the anomalous magnetic moment for a spin 1/2 fermion vanishes
in the SUSY limit[18] and away from the SUSY limit the cancellations
have somewhat less effect. Therefore, in general one can expect the Yukawa 
corrections from the  Higgs sector and
the supersymmetric electroweak 
corrections from virtual charginos and neutralinos  to cancel to some extent.

 In conclusion,   we calculated the Yukawa corrections of order 
$\alpha_{ew} M_t^2/M_W^2$ to  single
 top quark production via $q \bar q'\rightarrow t \bar b$ at
 the Fermilab Tevatron in the general two-Higgs-doublet model and 
the minimal supersymmetric model. 
For favorable parameter values,  the corrections can  
reduce the cross section by more than 15\% 
in the general two-Higgs-doublet model
and enhance the cross section by up to 10\% in the minimal supersymmetric model.
These effects may be observable at a high-luminosity Tevatron.

\vspace{.5cm}

This work was supported in part by the U.S. Department of Energy, Division
of High Energy Physics, under Grant No. DE-FG02-91-ER4086.

\vspace{.7cm}

\appendix{}

\begin{eqnarray} 
\delta Z_t^L&=&\frac{\alpha_{ew}}{16\pi M_W^2 s_W^2}
  \left\{ \sum_{i=H,h}M_t^2 \eta_i^2 [-\frac{\Delta}{2}+F_1^{(tti)}
	 +2 M_t^2 (G_0^{(tti)}+G_1^{(tti)})]\right.\nonumber\\
  & & +\sum_{i=A,G^0}M_t^2 \eta_i^2 [-\frac{\Delta}{2}+F_1^{(tti)}
	+2 M_t^2 (G_1^{(tti)}-G_0^{(tti)})]\nonumber\\
  & & \left. +\sum_{i=H^+,G^+}[2 M_b^2 \lambda_i^2 
(-\frac{\Delta}{2}+F_1^{(tbi)})
	+2 M_t^2 (M_t^2 \eta_i^2+M_b^2 \lambda_i^2)G_1^{(tbi)}]
						\right\}\\
\delta Z_b^L&=&\frac{\alpha_{ew}}{16\pi M_W^2 s_W^2}
  \left\{ \sum_{i=H,h} M_b^2 \lambda_i^2 [-\frac{\Delta}{2}+F_1^{(bbi)}
	+2 M_b^2 (G_0^{(bbi)}+G_1^{(bbi)}))]\right.\nonumber\\
  & & +\sum_{i=A,G^0}M_b^2 \lambda_i^2 [-\frac{\Delta}{2}+F_1^{(bbi)}
	+2 M_b^2 (G_1^{(bbi)}-G_0^{(bbi)})] \nonumber\\
  & & +\sum_{i=H^+,G^+}[ 2 M_t^2 \eta_i^2 (-\frac{\Delta}{2}+F_1^{(bti)})
	    + 4 M_b^2 M_t^2 \eta_i \lambda_i G_0^{(bti)}\nonumber\\
  & & \left.\hspace{1cm}
            +2 M_b^2 (M_t^2 \eta_i^2+M_b^2 \lambda_i^2) G_1^{(bti)}] \right\}
\end{eqnarray}
where $\Delta\equiv \frac{1}{\epsilon}-\gamma_E+\log 4\pi$
with $\gamma_E$ being the Euler constant and $D=4-2\epsilon$ is
the space-time dimension. 
The functions $F^{(ijk)}_n, G^{(ijk)}_n$ are defined by
\begin{eqnarray}
F^{(ijk)}_n&=&\int^1_0 dy y^n\log \left [\frac{M_i^2y(y-1)+M^2_j(1-y)
+M^2_k y}{\mu ^2}\right ],\\
G^{(ijk)}_n&=& -\int^1_0 dy \frac{y^{n+1}(1-y)}{M_i^2y(y-1)+
M^2_j(1-y)+M^2_ky}.
\end{eqnarray}

The form factors $f^L_{i,j}$ are obtained by
\begin{equation} 
f^{L,R}_i=\frac{\alpha_{ew}}{16\pi M_W^2 s_W^2}\sum_{j=1}^{6} f^{L,R}_{ij}
\end{equation} 
$f^{L,R}_{ij}$ are given by
\begin{eqnarray} 
f^L_{11}&=& 2 M_b^2\xi_{ij} \lambda_j \lambda_i \bar c_{24} \\
f^L_{21}&=& M_t M_b^2\xi_{ij} \lambda_j[\eta_i (2 c_{23}+c_{12}-2 c_{22})
 +\lambda_i (-c_{12}+2 c_{22}+2 c_{21}+c_{11}-4 c_{23})]\\ 
f^L_{31}&=& M_t M_b^2\xi_{ij} \lambda_j [\eta_i (-c_{12}-2 c_{22})
 +\lambda_i (-c_{11}+2 c_{22}-2 c_{23}+c_{12})]\\ 
f^R_{11}&=&2 M_t M_b \xi_{ij} \lambda_j\eta_i \bar c_{24} \\
f^R_{21}&=& M_b \xi_{ij} \lambda_j
		[M_b^2 \lambda_i (2 c_{23}+c_{12}-2 c_{22})\nonumber\\
 & &    +M_t^2 \eta_i (-c_{12}+2 c_{22}+2 c_{21}+c_{11}-4 c_{23})]\\
f^R_{31}&=& M_b \xi_{ij} \lambda_j
		[M_b^2 \lambda_i (-c_{12}-2 c_{22})
  	+M_t^2 \eta_i (-c_{11}+2 c_{22}-2 c_{23}+c_{12})]\\
f^L_{12}&=& 2 M_b^2 \xi_{ij} \lambda_j\lambda_i \bar c_{24} \\
f^L_{22}&=&- M_t M_b^2 \xi_{ij} \lambda_j
		[\eta_i (2 c_{23}+c_{12}-2 c_{22})\nonumber\\
 & &	-\lambda_i (-c_{12}+2 c_{22}+2 c_{21}+c_{11}-4 c_{23})] \\
f^L_{32}&=&- M_t M_b^2\xi_{ij} \lambda_j
		 [\eta_i (-c_{12}-2 c_{22})
	-\lambda_i (-c_{11}+2 c_{22}-2 c_{23}+c_{12})] \\
f^R_{12}&=&- 2 M_t M_b \xi_{ij} \lambda_j\eta_i \bar c_{24}\\ 
f^R_{22}&=& M_b\xi_{ij} \lambda_j
		 [M_b^2 \lambda_i (2 c_{23}+c_{12}-2 c_{22})\nonumber\\
& & 	-M_t^2 \eta_i (-c_{12}+2 c_{22}+2 c_{21}+c_{11}-4 c_{23})]\\
f^R_{32}&=& M_b \xi_{ij} \lambda_j
		[M_b^2 \lambda_i (-c_{12}-2 c_{22})
	-M_t^2 \eta_i (-c_{11}+2 c_{22}-2 c_{23}+c_{12})]\\
f^L_{13}&=&-2 M_t^2 \xi_{ij} \eta_j\eta_i \bar c_{24}\\ 
f^L_{23}&=&-M_t \xi_{ij} \eta_j
		[M_t^2 \eta_i (2 c_{21}+2 c_{22}-4 c_{23}
	+c_{12}-c_{11}-c_0)\nonumber\\
 & &  +M_b^2 \lambda_i (2 c_{23}-2 c_{22}+c_{12})]\\
f^L_{33}&=&-M_t\xi_{ij} \eta_j[M_t^2 \eta_i (2 c_{22}-2 c_{23}+3 c_{12}
	-c_{11}+c_0)-M_b^2 \lambda_i (2 c_{22}+c_{12})] \\
f^R_{13}&=&-2 M_t M_b \xi_{ij} \eta_j\lambda_i \bar c_{24}\\ 
f^R_{23}&=&-M_t^2 M_b \xi_{ij} \eta_j
		 [\eta_i (2 c_{23}-2 c_{22}+c_{12})\nonumber\\
 & & 	+\lambda_i (2 c_{21}+2 c_{22}-4 c_{23}+c_{12}-c_{11}-c_0)]\\
f^R_{33}&=&-M_t^2 M_b \xi_{ij} \eta_j [-\eta_i (2 c_{22}+c_{12})
	+\lambda_i (2 c_{22}-2 c_{23}+3 c_{12}-c_{11}+c_0)]\\
f^L_{14}&=&2 M_t^2 \xi_{ij} \eta_j\eta_i \bar c_{24}\\ 
f^L_{24}&=&-M_t\xi_{ij} \eta_j
	 (M_t^2 \eta_i (4 c_{23}-2 c_{21}-2 c_{22}
	+3 c_{12}-3 c_{11}-c_0)\nonumber\\
 	& &+M_b^2 \lambda_i (2 c_{23}-2 c_{22}+c_{12})] \\
f^L_{34}&=&- M_t \xi_{ij} \eta_j
		[M_t^2 \eta_i (2 c_{23}-2 c_{22}+c_{12}
	+c_{11}+c_0)-M_b^2 \lambda_i (2 c_{22}+c_{12})]\\ 
f^R_{14}&=&- 2 M_t M_b \xi_{ij} \eta_j\lambda_i \bar c_{24} \\
f^R_{24}&=&-M_t^2 M_b \xi_{ij} \eta_j
		[\eta_i (2 c_{22}-2 c_{23}-c_{12})\nonumber\\
 & &+\lambda_i (2 c_{21}+2 c_{22}-4 c_{23}+3 c_{11}-3 c_{12}+c_0)]\\
f^R_{34}&=&- M_t^2 M_b \xi_{ij} \eta_j
		[\eta_i (2 c_{22}+c_{12})
	+\lambda_i (2 c_{22}-2 c_{23}-c_{12}-c_{11}-c_0)]\\
f^L_{15}&=&-M_t^2 M_b^2\eta_j \lambda_j (c_0+2 c_{12}-c_{11}) \\
f^L_{25}&=&2 M_t M_b^2 \eta_j \lambda_j(c_{12}+c_{23})\\
f^L_{35}&=&2 M_t M_b^2 \eta_j \lambda_j(c_{12}+c_{22})\\
f^R_{15}&=&M_t M_b \eta_j \lambda_j[2 \bar c_{24}+\hat s (c_{12}+c_{23})
	-M_t^2 (c_0-c_{21}+c_{12}+c_{23})] \\
f^R_{25}&=&2 M_t^2 M_b \eta_j \lambda_j(c_0-c_{21})\\
f^R_{35}&=&2 M_t^2 M_b \eta_j \lambda_j(c_{12}+c_{23})\\
f^L_{16}&=&M_t^2 M_b^2 (c_0+c_{11}) \\
f^L_{26}&=&2M_t  M_b^2 (c_{12}+c_{23})\\
f^L_{36}&=&2M_t M_b^2 (c_{12}+c_{22})\\
f^R_{16}&=&M_t M_b [2 \bar c_{24}+\hat s (c_{12}+c_{23})
	-M_t^2 (c_{12}+c_{23}-c_{21}-2 c_{11}-c_0)]\\ 
f^R_{26}&=&- 2 M_t^2 M_b (c_0+2 c_{11}+c_{21})\\
f^R_{36}&=&2 M_t^2 M_b (c_{12}+c_{23})\\
\end{eqnarray} 
where the sums over $i=H^+,G^+, j=H,h$ for $f^{L,R}_{i1}$ and $f^{L,R}_{i3}$, 
$i=H^+,G^+, j=A,G^0$ for $f^{L,R}_{i2}$ and $f^{L,R}_{i4}$, and 
$j=H,h$ for $f^{L,R}_{i5}$ are implied. The functions $c_{ij}$ defined as
\begin{eqnarray*} 
c_{ij}&=&c_{ij}(-p_3,k,M_b,M_i,M_j) {~~\rm for~~}  f^{L,R}_{i1}, f^{L,R}_{i2}\\
c_{ij}&=&c_{ij}(-p_3,k,M_t,M_j,M_i) {~~\rm for~~}  f^{L,R}_{i3}, f^{L,R}_{i4}\\
c_{ij}&=&c_{ij}(-p_3,-p_4,M_t,M_j,M_b) {~~\rm for~~}  f^{L,R}_{i5}\\
c_{ij}&=&c_{ij}(-p_3,-p_4,M_t,M_A,M_b) {~~\rm for~~}  f^{L,R}_{i6}
\end{eqnarray*} 
are the three-point Feynman integrals[19].
The constants $\eta_i, \lambda_i$ and $\xi_{ij}$ are defined as
\begin{eqnarray} 
        \eta_{H^+}&=&\eta_A=\cot\beta,~~
        \eta_{G^+}=\eta_{G^0}=1,\\
        \eta_H&=&\sin\alpha/\sin\beta,~~
        \eta_h=\cos\alpha/\sin\beta,\\
        \lambda_{H^+}&=&\lambda_A=\tan\beta,~~
        \lambda_{G^+}=\lambda_{G^0}=0,\\
        \lambda_H&=&\cos\alpha/\cos\beta,~~
        \lambda_h=-\sin\alpha/\cos\beta,\\
	\xi_{H^+H}&=&-\xi_{G^+h}=\sin(\beta-\alpha),\\
	\xi_{H^+h}&=&\xi_{G^+H}=-\cos(\beta-\alpha),\\
 	\xi_{H^+A}&=&\xi_{G^+G^0}=1,~~
	\xi_{H^+G^0}=\xi_{G^+A}=0
\end{eqnarray}
\eject
 
{\LARGE References}
\vspace{0.3in}
\begin{itemize}
\begin{description}
\item[{\rm [1]}] CDF Collaboration,  Phys.Rev.Lett. {\bf 74}, 2626(1995);\\
                D0 Collaboration,  Phys.Rev.Lett. {\bf74}, 2632(1995).
\item[{\rm[2]}] F.Berends, J.Tausk and W.Giele, Phys.Rev.D47, 2746(1993).
\item[{\rm[3]}] S.Willenbrock and D.Dicus, Phys.Rev.D34, 155(1986);\\
		S.Dawson and S.Willenbrock, Nucl.Phys.B284, 449(1987);\\
		C.P.Yuan, Phys.Rev.D41, 42(1990);\\
	       	F.Anselmo, B.van Eijk and G.Bordes, Phys.Rev.D45, 2312(1992);\\
		R.K.Ellis and S.Parke, Phys.Rev.D46,3785(1992);\\
		D.Carlson and C.P.Yuan, Phys.Lett.B306,386(1993);\\
         	G.Bordes and B.van Eijk, Nucl.Phys.B435, 23(1995);\\
		A.Heinson, A.Belyaev and E.Boos, hep-ph/9509274.
\item[{\rm[4]}] S.Cortese and R.Petronzio, Phys.Lett.B306, 386(1993).
\item[{\rm[5]}] T.Stelzer and S.Willenbrock, Phys.Lett.B357, 125(1995).
\item[{\rm[6]}] M.Smith and S.Willenbrock, hep-ph/9604223. 
\item[{\rm[7]}] For a review, see, for example, J. F. Gunion et. al.,
                {\bf The Higgs Hunters' Guide} (Addison-wesley, Teading,
 		MA, 1990).
\item[{\rm[8]}] E.Farhi and L.Susskind, Phys.Rev.D20, 3404(1979);\\ 
		J.Ellis et al., Nucl.Phys.B182, 529(1981);\\
		C.D.Carone and H.Georgi, Phys.Rev.D49, 1427(1994).
\item[{\rm[9]}] A. Sirlin, Phys. Rev. D22(1980)971;\\
                W.J. Marciano and A. Sirlin, {\it ibid.} 22, 2695(1980);
                                             31,213(E)(1985);\\
                A. Sirlin and W.J. Marciano, Nucl.Phys.B189(1981)442;\\        
                K.I.Aoki et al., Prog.Theor.Phys.Suppl. 73(1982)1.
\item[{\rm[10]}] H.L. Lai et.al., Phys.Rev.D51, 4763(1995).
\item[{\rm[11]}] A Stange and S.Willenbrock, Phys.Rev.D48,2054(1993).
\item[{\rm[12]}] A.D. Martin, R.G. Roberts and W.J. Stirling,
		 Phys. Lett. B354, 155(1995).
\item[{\rm[13]}] For example, see F.Zwirner, CERN-TH.6357/91;\\
                 V. Barger, K. Cheung, R.J.N.Phillips and A.L.Stange, 
		Mad/PH/704;\\
                Z.Kunszt and F.Zwirner, Nucl.Phys.B385(1992)3.
\item[{\rm[14]}] OPAL Coll. CERN-PPE/95-180.
\item[{\rm[15]}] ALEPH Coll. CERN-PPE/94-165.
\item[{\rm[16]}] V.Barger, M.S.Berger and P.Ohmann, Phys.Rev.D47, 1093(1993).
\item[{\rm[17]}] C.S.Li, R.J.Oakes and J.M.Yang, work in progress.
\item[{\rm[18]}] S.Ferrara and E.Remiddi, Phys.Lett.B53(1974)347;\\
                 P. Fayet and S. Ferrara, Phys. Rep. C32 (1977) No.5.
\item[{\rm[19]}] G. Passarino and M. Veltman, Nucl. Phys. B160(1979)151.

\end{description}
\end{itemize}
\vfil
\eject

\begin{center} {\bf Figure Captions} \end{center}
\vspace{.7cm}

Fig.1 Feynman diagrams: (a) tree-level, (b)-(e) self-energies,
 (f)-(h) vertex corrections.

Fig.2 The relative correction $\Delta\sigma/\sigma_0$  as a function of 
$M_H$ using the CTEQ3L parton distributions for $\sigma_0$ and the CTEQ3M 
parton distributions for $\Delta\sigma$.
The solid curve corresponds to the 2HDM 
assuming $M_h=M_H, M_t=175$ GeV, $M_{H^+}=M_A=600$GeV and $\tan\beta=0.25$.
The dotted curve corresponds to the SM for $M_t=175$ GeV.

Fig.3 The relative correction $\Delta\sigma/\sigma_0$  as a function of 
$M_{H^+}$ in the MSSM, assuming $\tan\beta=0.25$.
The solid curve corresponds to $M_t=175$ GeV using the 
CTEQ3L parton distributions for $\sigma_0$ and the CTEQ3M parton distributions
for $\Delta\sigma$.
The dotted curve corresponds to $M_t=175$ GeV using the MRSG parton 
distributions.
The dashed curve corresponds to $M_t=200$ GeV using 
the CTEQ3L distributions for $\sigma_0$ and the CTEQ3M distributions 
for $\Delta\sigma$.

Fig.4 The relative correction $\Delta\sigma/\sigma_0$  as a function of 
$\tan\beta$ using the CTEQ3L distributions 
for $\sigma_0$ and the CTEQ3M distributions for $\Delta\sigma$. 
The solid curve corresponds to the 2HDM
assuming $M_t=175$GeV, $M_{H^+}=M_A=600$GeV and $M_H=M_h=65$GeV.
The dotted curve corresponds to the MSSM 
assuming $M_t=175$GeV and $M_{H^+}=100$GeV.

\end{document}